\documentstyle[twocolumn,pra,aps]{revtex}
\begin{document}
\title{Bose-Einstein condensation with magnetic dipole-dipole forces}
\author{Krzysztof G\'{o}ral$\dagger$, Kazimierz
Rz\c{a}\.{z}ewski$\dagger$ and Tilman Pfau$\S ^{\flat}$}
\address{$\dagger$ Center for Theoretical Physics and College of Science,
Polish Academy of Sciences,\\ Aleja Lotnik\'ow 32/46, 02-668 Warsaw,
Poland\\ 
${\S}$ Faculty of Physics, University of Konstanz, 78457 Konstanz,
Germany }

\date{\today}

\maketitle

\begin{abstract}
Ground-state solutions in a dilute gas interacting via contact and
magnetic dipole-dipole forces are investigated. To the best of our
knowledge, it is the first example of studies of the Bose-Einstein
condensation in a system with realistic long-range interactions.
We find that for the magnetic moment of e.g. chromium (6$\mu_B$)
and a typical value of the scattering length all solutions are
stable and only differ in size from condensates without long-range
interactions. By lowering the value of the scattering length we
find a region of unstable solutions. In the neighborhood of this
region the ground state wavefunctions show internal structures not
seen before in condensates. Finally, we find an analytic estimate
for the characteristic length appearing in these solutions.
\end{abstract}
\pacs{PACS numbers: 03.75.Fi, 05.30.Jp}

Since the advent of Bose-Einstein condensation in dilute gases of
alkalies \cite{BEC_exp} and hydrogen \cite{H_exp} it has become
apparent that the interactions between the condensed atoms govern
most of the observed phenomena. In all the experiments so far the
interaction can be described by a contact potential which is
characterized by the scalar quantity $a$ being the s-wave
scattering length. Static properties like the condensate's
ground-state density profile, its instability in the case of
negative $a$ and the mean-field shift in spectroscopic
measurements as well as dynamic properties like collective
excitations and propagation of sound have been investigated
\cite{varenna}. Similarly, nonlinear atom optics experiments e.g.
four wave mixing \cite{fourwave} are only possible due to the
large nonlinearity mediated by the atom-atom interactions. The
$T$=0K situation in almost all experiments can be described very
well by the Gross-Pitaevskii equation \cite{GP}.

Any reasonably strong dipole--dipole interaction would largely
enrich the variety of phenomena to be observed in dilute gases due
to their long range and vectorial character. However, for all of
the condensed atomic species the magnetic moment $\mu$ was roughly
1 $\mu_B$ (Bohr magneton) and the respective magnetic
dipole--dipole interaction was negligible compared to the contact
potential. It has been proposed to induce a strong electric
dipole--dipole interaction in alkalies by the application of
strong DC electric fields \cite{marinescu}.

Recently it has become possible to trap atoms with higher magnetic
moments at high densities. Examples are europium ($\mu = 7 \mu_B)$
\cite{buffer_eu}, which has been trapped magnetically by a
buffer-gas loading technique, and chromium ($\mu = 6 \mu_B$),
which has been loaded into a magnetic trap by a buffer-gas
technique \cite{buffer_cr} and by laser cooling \cite{LC_cr}. For
these species the scattering lengths are not known to date, but,
assuming a normal non-resonant behavior, the crossections of the
scalar contact potential and the magnetic dipole--dipole
interaction are of comparable size.

As the dipole--dipole interaction is attractive parallel to a
common polarization axis the immediate question arises: can a
stable condensate be formed under the influence of a
dipole--dipole interaction? What is its effect on anisotropic
clouds? What do the ground-state wavefunctions look like? In this
Letter we address these questions and choose the magnetic
dipole--dipole interaction in a dilute cloud of chromium atoms as
an example for numerical calculations. The results however apply
for all static dipole-dipole interactions.

In general, the Gross-Pitaevskii equation with a long-range
interaction term and for a cylindrical harmonic trap has the
following form:
\begin{eqnarray} \label{GP}
i\hbar\frac{\partial \Psi}{\partial t}= \{
-\frac{\hbar^{2}\triangle}{2m} +\frac{1}{2} m \omega_{0}^{2}(x^{2}+
y^{2}+\gamma^{2}z^{2})&+&\\ +\frac{4 \pi
\hbar^{2}a}{m}N|\Psi|^{2}\mbox{}+ N \int
V(\vec{r}-\vec{r'})|\Psi(\vec{r'})|^{2}d^{3}\vec{r'} \} \Psi& &
\nonumber ,
\end{eqnarray}
where $\Psi$ is the mean-field condensate wavefunction, $a$ -- the
s-wave scattering length, $N$ -- number of atoms and $m$ -- mass
of the atom. The reference frequency of the trap $\omega_{0}$ is
chosen in the $xy$ plane and the anisotropy of the trap is defined
by the $\gamma$ factor. In our case, the long-range potential
takes the form characteristic of magnetic dipole-dipole
interactions:
\begin{equation}
        V(\vec{r}-\vec{r'})=\frac{\mu_{0}}{4 \pi}\;
        \frac{ \vec{\mu_{1}}(\vec{r}) \cdot \vec{\mu_{2}}(\vec{r'}) -
3\vec{\mu_{1}}(\vec{r})\cdot\vec{u} \;
\vec{\mu_{2}}(\vec{r'})\cdot\vec{u}}
        {|\vec{r}-\vec{r'}|^3},
\end{equation}
where $\vec{u}=\frac{\vec{r}-\vec{r'}}{|\vec{r}-\vec{r'}|}$ and
$\mu_{0}$ is magnetic permeability of the vacuum. We will assume
that all the magnetic moments are in the same direction
($\vec{\mu_{1}}=\vec{\mu_{2}}$), which will be referred to as the
polarization direction. Note that, depending on a configuration of
dipoles, magnetic potential can be repulsive as well as
attractive. This fact is a source of a variety of phenomena which
do not appear in condensates with contact interactions only.
Another peculiar feature of this long-range (integral) term is
that for a uniform density distribution it vanishes if integrated
within a sphere, which means that in that case a dipole placed in
the middle of the sphere would feel no magnetic force.

Equation (\ref{GP}) is now an integro-differential equation. The
integral term can be simplified, because a part of it can be
calculated analytically. First of all, one should notice that this
term has a form of a convolution:
\begin{equation}
\int V(\vec{r}-\vec{r'})|\Psi(\vec{r'})|^{2}d^{3}\vec{r'}=
V(\vec{r}) \ast |\Psi(\vec{r})|^{2}
\end{equation}
The transform of the potential reads:
\begin{equation}
{\cal F}(V(\vec{r}))=\mu_{0} \mu^{2}\;(1- 3\cos^{2} \alpha)\; [
\frac{\cos(bq)}{(bq)^{2}} - \frac{\sin(bq)}{(bq)^{3}} ],
\end{equation}
where $b$ is a distance below which atoms overlap (i.e. the radius
of the atoms which is of the order of a few Bohr radii) and
$\alpha$ is the angle between the Fourier variable $\vec{q}$ and
$\vec{\mu}$. A value of $b$ is small in comparison with a length
scale in the system (oscillator unit) and we will not consider
large values of $q$ therefore it is sufficient to use the limit:
\begin{equation}
\lim_{b \rightarrow 0} {\cal F}(V(\vec{r})) = - \frac{1}{3}
\mu_{0} \mu^{2}\;(1 - 3\cos^{2} \alpha).
\end{equation}
In order to evaluate the Fourier transform of  eq. (3) ${\cal F}(|
\psi|^{2})$ is calculated numerically (FFT) and multiplied by
${\cal F}(V(\vec{r}))$which depends solely on $\alpha$.

In order to obtain the ground-state solution for the condensate
one has to solve the Gross-Pitaevskii equation in the form of
Eq.(\ref{GP}). To do it  we employ a variant of the split-operator
method called the imaginary-time propagation method, being now a
common routine of solving the Gross-Pitaevskii equation. For each
time-step one needs to compute four FFT's: two for the long-range
term calculation and two for the evolution.

Now we will list parameters used in the calculations. The
reference trap frequency $\omega_{0}$ is equal to $2 \pi\, 150$~Hz
which is an accessible and typical value in present experiments.
This corresponds to a characteristic length unit in the system
(oscillator unit) $d=\sqrt{\frac{\hbar}{m_{Cr} \omega_{0}}}
\simeq$~1.14$~\mu$m. The s-wave scattering length for chromium is
unknown -- instead a value for sodium is assumed tentatively
$a=a_{Na}=$ 2.75 nm.

One should keep in mind that all the presented solutions are
scalable. For the Gross-Pitaevskii equation with as well as
without long-range forces the solutions stay the same -- in scaled
coordinates -- as long as the product $N\sqrt{\omega_{0}}$ is kept
constant.

The ratio of the contact and long-range terms in general is given
by $\frac{\mu^2 m}{a}$, $m$ being the atom's mass, and solutions
stay identical if it is kept constant. For the chosen example of
chromium we will in the second part of the Letter vary the
scattering length to change that ratio.

The main conclusion is that, for this set of parameters (in
particular the scattering length value) all the investigated
solutions are stable, even in very asymmetric traps. Moreover, the
wavefunctions possess familiar shapes: Gaussian-like for a small
number of atoms and parabolic-like for bigger condensates.

At first, we investigated two extreme cases in cylindrical traps.
Following simple intuition about an interaction of two dipoles, we
designed these situations so that in the first one there would be
a majority of repulsive forces, whereas in the other one an
attractive component would be dominant.

In the first situation, the trap is flattened in a plane
perpendicular to the polarization axis ($xy$ plane) with the
asymmetry factor $\gamma=\frac{\omega_z}{\omega_0}=10$ (referred
to as disc-shaped). For this, and only this case, we lower the
reference frequency to $\omega_{0}=2 \pi\, 15$~Hz to make
$\omega_z=10 \omega_{0}$ experimentally accessible (the
corresponding oscillator unit is $d=\simeq$~3.6$~\mu$m). In the
inset of Fig.\ref{F1} half-width of the ground state vs. number of atoms
is plotted for the disc-shaped trap. It is compared to a
corresponding half-width for a Thomas-Fermi solution for a
condensate with contact interactions only (referred to as the
standard Thomas-Fermi limit). Note that the Thomas-Fermi
approximation is good only for large number of atoms, thus for
$N<$ 10,000 it has not been used. In the $z$-direction the exact
solution is indistinguishable from the standard Thomas-Fermi
limit, because $z$ is a stiff direction and so all the
interparticle interaction effects are small with respect to the
trapping forces. On the other hand, along the $x$-axis the
condensate expands with respect to the standard Thomas-Fermi limit
as it should be the case for the repulsive direction. To conclude,
in the disc-shaped trap the net effect of interactions is
repulsive and the condensate expands.

The opposite extreme case is a trap stretched along the
polarization axis ($z$-axis) with $\gamma=0.1$ (referred to as
cigar-shaped) -- see Fig.\ref{F1}. For the soft direction (along
the $z$-axis, being also an attractive direction), in agreement
with simple intuition, one observes shrinking of the condensate.
The differences with respect to the standard Thomas-Fermi limit
for a condensate without long-range interactions diminish as the
number of atoms grows. This is the case because in the $z$
direction the wavefunction is very flat (roughly uniform) and so
the integral component almost vanishes. Surprisingly, we observe
shrinking of the condensate in the $x$ (repulsive) direction as
well. We argue that it is caused by the attraction exerted by very
many dipoles concentrated along the soft ($z$) direction which
takes over the repulsive contribution along the $x$-axis.
Concluding, in the cigar-shaped trap we observe an overall
shrinking of the condensate.

As we see, long-range magnetic interactions do not change
qualitatively the ground-state solutions of the Gross-Pitaevskii
equation, but they affect mainly a size of the condensate. This
indicates that for the used set of parameters an influence of
contact (repulsive) interactions is still dominant.  However, as we do not
know the scattering length for chromium and there are several methods
proposed to manipulate this parameter by
application of various external fields (e.g. see \cite{modscatt},
\cite{fedichev}), we decreased the assumed value and by this means
enhanced the effect of long-range (partially attractive) forces.
As a result, we were able to find a region of unstable solutions.
An even more striking discovery was the observation of structured
shapes of the wavefunctions acquired by the condensates near the
instability threshold. By probing the parameter space $(a,N)$
(scattering length, number of atoms) we found the phase diagram
depicting localizations of stable, structured (still stable) and
unstable solutions shown in Fig.\ref{F2}. For simplicity we did
calculations for a spherical trap with a frequency
$\omega_0$=$2\pi\, 150$~Hz, but a few cases calculated for
asymmetric traps convinced us that the qualitative stability
behavior would not change. The polarization direction is still
$z$. For illustration purposes, we present two examples of the
structured solutions. The first one (Fig.\ref{F3}) was obtained
for 80,000 atoms with $a/a_{Na}$=0.115 which situates this case
very near the instability threshold. The second example
(Fig.\ref{F4}) is a wavefunction for 4,000,000 atoms and
$a/a_{Na}$=0.233 (again right above the instability threshold). We
have also used geometry of the trap to stabilize solutions
unstable in a spherically symmetric potential (by increasing the
asymmetry factor $\gamma$) as well as to destabilize ones that
were stable in a spherical trap (by decreasing $\gamma$). It is remarkable
that the complexity of the ground-state wavefunction arises solely
from atom-atom interactions and does not reflect simplicty of an external
potential (it represents self-organization in the ground state).

For all the structured solutions there exists a characteristic
length defined by a distance between adjacent maxima in the plane
perpendicular to polarization, which is roughly 3$d$. In order to
explain this feature, we performed a stability analysis for an
infinite (no trap), homogenous case. In this case a proper
solution is $\Psi=\frac{1}{\sqrt{V}}\exp(-igt)$, $V$ being the
volume and $g=N \frac{4 \pi \hbar a}{m_{Cr}}$. By imposing a small
harmonic perturbation of the
form $\Psi=\frac{1}{\sqrt{V}}[1+u(t) \cos(\vec{q} \cdot
\vec{r})]\exp(-igt)$ we found that the unstable perturbations are
those for which:
\begin{equation}
q^2<\frac{4Nm_{Cr}}{\hbar
V}[\frac{\mu_{0}\mu^{2}}{3\hbar}(1-3\cos^2\alpha)-\frac{4 \pi
\hbar a}{m_{Cr}}],
\end{equation}
where $\alpha$ is the angle between the wavevector and the
polarization direction. This result implies that surprisingly all
the perturbations parallel to polarization ($\alpha'=0$) are
stable, whereas for the direction perpendicular to polarization
($\alpha=\frac{\pi}{2}$) there is a long-wave instability. To
compare this result with the observed characteristic length of
3$d$ we crudely approximated the volume of our condensates by
$V=\frac{4}{3} \pi R_{TF}^{3}$, where $R_{TF}$ is the Thomas-Fermi
radius for a condensate without long-range interactions. This
yields the critical wavelength:
\begin{equation}
\lambda_{C}^{2}=\frac{4 \pi^{3} \hbar}{3 N^{\frac{2}{5}} m_{Cr}}
\frac{(15a
)^{\frac{3}{5}}d^{\frac{12}{5}}}{[\frac{\mu_{0}\mu^{2}}{3\hbar}-\frac{4
\pi \hbar a}{m_{Cr}}]}.
\end{equation}
For the two structured wavefunctions presented in this letter,
this value corresponds to 2.2$d$ and 1.8$d$, respectively, which,
for such an approximate analysis, is in a surprisingly good
agreement. We note in passing that the onset of unstable
perturbations ($\lambda_{C}\rightarrow\infty$) corresponds to
$a_{C}\simeq 0.3 a_{Na}$ which, again, roughly approximates the
calculated instability threshold (see Fig.\ref{F2}).

Motivated by  experiments under development, all our calculations used
chromium parameters. They showed that ground-state solutions differ in a
non-trivial way from the usual solutions and found a region of instability
as well as structured solutions in its neighborhood. Reaching the
interesting region near the instability boundary for chromium seems
problematic with state-of-the-art techniques. However, in the case of
europium \cite{buffer_eu} (larger mass and magnetic moment) the
instability threshold for 1,000,000 atoms is situated at 92\% of the
sodium
scattering length value and still grows for larger numbers of atoms (e.g.
108\% for 10,000,000 atoms). Our results may also find use in the blooming
area of cold molecules \cite{CaH} some of which possess large permanent
electric dipole moments \cite{slowing}. For a typical electric dipole
moment value of 1 Debye a prefactor of the long-range term is 4 orders of
magnitude greater than the corresponding term for magnetic dipole of 1
Bohr magneton. One can thus conclude that behavior of a polar molecular
BEC will be overruled by dipole forces.

We would like to thank G.V.Shlyapnikov, M.Brewczyk, Z.Idziaszek
and M.Gajda for helpful discussions. K.G. acknowledges support by
Polish KBN grant no 203B05715. K.R. and K.G. are supported by the
subsidy of the Foundation for Polish Science. T.P. and K.R.
acknowledge support by the Alexander von Humboldt foundation. The
results have been obtained using computers at the
Interdisciplinary Centre for Mathematical and Computational
Modeling (ICM) at Warsaw University.

{\it $\flat$:  current adress: University of Stuttgart, Physics
Department, 5th Institute, D-70550 Stuttgart, Germany.}
\begin{figure}
\caption{Half-width of the chromium condensate vs. number of atoms
in a cigar-shaped cylindrical trap with $\gamma=0.1$ and $a$=$a_{Na}$.
Filled circles indicate an exact solution for a section along the $x$-axis
as compared to the corresponding result in the standard Thomas-Fermi limit
(empty circles). Empty squares represent an exact solution along the
$z$-axis in comparison with the corresponding standard Thomas-Fermi limit
(filled squares). The inset presents analogous results in a disc-shaped
trap ($\gamma=10$) and for $a$=$a_{Na}$. An exact solution for a section
along the $z$-axis (filled  squares) is indistinguishable from the
standard Thomas-Fermi limit. \label{F1}}
\end{figure}
\begin{figure}
\caption{Stability diagram for a spherical trap with
$\omega_0$=$2\pi\, 150$~Hz. Number of atoms is plotted  on the
horizontal axis and the scattering length (expressed as a fraction
of the corresponding value for sodium) on the vertical one. Solid
line is the instability threshold -- below it no stable solutions
could be obtained. The dashed line is the boundary between
standard-shaped and structured solutions. In the shaded area
structured stable solutions are present. Error bars indicate
discreteness of our parameter-space probing. \label{F2}}
\end{figure}
\begin{figure}
\caption{Squared modulus of the ground-state wavefunction in the
$xz$ plane for a spherical trap with $\omega_0$=$2\pi\, 150$~Hz,
80,000 atoms and the scattering length $a/a_{Na}$=0.115. The
horizontal axes are in oscillator units $d$. \label{F3}}
\end{figure}
\begin{figure}
\caption{Squared modulus of the ground-state wavefunction in the
$xz$ plane for a spherical trap with $\omega_0$=$2\pi\, 150$~Hz,
4,000,000 atoms and the scattering length $a/a_{Na}$=0.233. The
horizontal axes are in oscillator units $d$. \label{F4}}
\end{figure}

\end{document}